# Bias-driven large power microwave emission from MgO-based tunnel magnetoresistance devices

Alina M. Deac[2,1], Akio Fukushima[1], Hitoshi Kubota[1], Hiroki Maehara[3,2,1], Yoshishige Suzuki[2,1], Shinji Yuasa[1], Yoshinori Nagamine[3], Koji Tsunekawa[3], David D. Djayaprawira[3] and Naoki Watanabe[3]

[1]*National Institute of Advanced Industrial Science and Technology (AIST), Nanoelectronics Research Institute, Tsukuba, Japan*
[2]*Osaka University, Graduate School of Engineering Science, Osaka, Japan*
[3]*Canon ANELVA Corporation, Electron Device Division, Kawasaki, Japan*

**Spin-momentum transfer between a spin-polarized current and a ferromagnetic layer can induce steady-state magnetization precession, and has recently been proposed as a working principle for ubiquitous radio-frequency devices for radar and telecommunication applications. However, to-date, the development of industrially attractive prototypes has been hampered by the inability to identify systems which can provide enough power. Here, we demonstrate that microwave signals with device-compatible output power levels can be generated from a single magnetic tunnel junction with a lateral size of 100 nm, seven orders of magnitude smaller than conventional radio-frequency oscillators. We find that in MgO magnetic tunnel junctions the perpendicular torque induced by the spin-polarized current on the local magnetization can reach 25% of the in-plane spin-torque term, while exhibiting a different bias-dependence. Both findings contrast with the results obtained on all-metallic structures - previously investigated -, reflecting the fundamentally different transport mechanisms in the two types of structures.**

A spin-polarized current flowing through a ferromagnet exerts a torque on the local magnetization and can induce steady-state precession under certain conditions[1,2]. This phenomenon has been intensively investigated in all-metallic giant magnetoresistance (GMR) multilayers using various configurations and experimental techniques[3-11]. It has been shown, for example, that the precession frequency can be adjusted by changing the current[11], and consequently spin-transfer induced precession has been proposed as working principle for new radio-frequency (RF) devices – such as frequency-tuneable microwave sources and resonators, nanometre scale transmitters and receivers, signal mixers, signal amplifiers and so on. However, even in the best cases, the output levels of such structures remain below 1 nW, far too low for applications, which require levels in the μW range[7]. It has been suggested that an array of phase-locked metallic oscillators could provide enough power, but phase-locking more than two devices[7,8] remains technologically difficult and yet unproven. Alternatively, MgO-based magnetic tunnel junctions (MTJs)[12,13] would inherently generate large signals due to their high tunnelling magnetoresistance (TMR) - above 400% at room temperature[14,15] -, but the physics of spin-momentum transfer in such structures is still to be elucidated. Indeed, recent theoretical models suggested that spin-transfer effects in MTJs might exhibit different trends compared to metallic pillars, in terms of bias-voltage-dependence and relative importance of the perpendicular ($T_\perp$) and in-plane ($T$) spin-torque terms, where the torque direction is given with respect to the plane defined by the magnetic moments of the free and reference layers[16,17]. In MTJs, the ratio between $T$ and $T_\perp$ and their bias dependence are influenced by the exchange splitting[17], the type/distribution of interfacial defects and/or the state density dependence on energy[16]. These predictions have been partially confirmed by means of spin-torque-driven ferromagnetic resonance measurements (ST-FMR)[18,19], where a small ac current with GHz frequency drives the magnetization of the free layer to resonance, thereby inducing a finite output dc voltage. By super-imposing an additional dc bias, it is possible to estimate the spin-torque bias dependence. However, such measurements offer little insight on the characteristics of steady-state large-angle precession driven by spin-transfer. Details such as the output power levels and their dependence on the applied bias are important issues for oscillator applications. Indeed, to-date, a quantitative study of spin-transfer driven dynamics in MgO-MTJs is still missing[20].

Here, we present results obtained by performing a different experiment. We apply (only) a dc bias in order to excite precession and we monitor the RF output voltage to probe the state of the free layer of an MgO-MTJ. We show that in the spin-torque driven precession (STP) regime, MgO-MTJs generate output powers in the µW range, compatible with applications for mobile telecommunications. We present a complete analysis of spin-transfer induced dynamics over a large range of applied currents and in-plane fields. We demonstrate that the magnitude of the perpendicular torque generated by the spin polarized current on the free layer of a MgO-MTJ can reach 25% of that of the in-plane spin-torque term, in contrast with the all-metallic samples where the perpendicular torque is negligible[21,22]. We find that, at low voltages, the perpendicular torque exhibits a quadratic bias dependence, while the in-plane term varies linearly with the applied bias, thus partially confirming both theoretical predictions[16,17] and the results of ST-FMR experiments[18,19]. The higher bias behaviour of both spin-torque terms differs from thoretical excpectations and is yet to be understood.

Our samples were nanopillars with $70 \times 160$ nm$^2$ elliptical cross-section, patterned from sputtered multilayers with the following structure: SiO$_2$ substrate / buffer layer / Pt$_{50}$Mn$_{50}$ 15 / Co$_{70}$Fe$_{30}$ 2.5 / Ru 0.85 / Co$_{60}$Fe$_{20}$B$_{20}$ 3 / Mg 0.6 / MgO 0.6 / Co$_{60}$Fe$_{20}$B$_{20}$ 1.5 / capping layer (thickness in nm) (Fig. 1a). The ultra-thin free layer was chosen so as to minimize the switching currents/voltages, which are proportional to the layer's thickness. In addition, the saturation magnetization ($M_s$) of the 1.5 nm CoFeB free layer was low (880 emu/cm$^3$)[23,24], further contributing to the reduction of the critical currents. Previous studies indicated that such structures have lower thermal stability than samples with thicker free layers (≥ 2 nm) and higher $M_s$[23,24]. It is to be noted that the 3 nm thick reference layer required higher bias for being excited and thus remained stationary within most of the investigated current range.

Fig. 1b presents a magnetoresistance loop corresponding to the switching of the free layer between the low-resistance parallel (P) state and the high-resistance antiparallel (AP) alignment. The TMR and the resistance-area (*RA*) product in the parallel state were 110% and 4 Ω·µm$^2$, respectively. The coercivity of the free layer was low (6 Oe). A small loop shift was measured, so that in zero applied fields, the sample was in the AP state. This shift is the average result of ferromagnetic coupling through barrier roughness and antiferromagnetic dipolar interaction between layers. The exchange-biased synthetic antiferromagnetic reference layer starts to rotate around 700 Oe. The switching currents for such samples were around ±0.3 mA (Supplementary Methods).

Microwave measurements were performed at room temperature between 0.7 and 20 GHz on a setup shown schematically in Fig. 1a. The current was varied between -1.1 and 1.1 mA with 0.05 mA steps. Fields with values from -900 to 900 Oe (increased with 50 Oe steps) were applied along the easy axis of the ellipse. Positive current is defined as electrons flowing from the free to the reference layer, thus favouring the AP state. Positive fields oppose the magnetization of the reference layer. A 40 dB preamplifier was used, but both the amplification and the background noise have been subtracted from the data presented here.

Fig. 2 shows spectra measured at intermediate positive and negative fields for different currents. At 200 Oe and positive current (Fig. 2b), both current and field favour the AP state. Because the in-plane spin-torque acts to increase the damping, it cannot drive the magnetization motion and only low power thermally excited ferromagnetic resonance (TE-FMR) signals are obtained. The spectra present two peaks centred around 3.5 and 5 GHz and are only slightly altered by changing the current. The precession frequency calculated for an external field of 200 Oe – using the Kittel formula[25], the measured $M_s$ and neglecting the coupling fields – is about 4.3 GHz (Supplementary Discussion). The two peaks fall on each side of the calculated value (Supplementary Figure 5). We tentatively attribute the lower frequency peak to a centre precession mode, since the dipolar field from the reference layer is low at the centre of the ellipse and the ferromagnetic coupling through barrier roughness opposes the external field, thus reducing the total field acting on the local magnetization. The higher frequency peak likely arises from a mode with large amplitude at the two ends of the long axis of the ellipse, where the antiferromagnetic magnetostatic interaction between the layers is strong and possibly the local field is higher than the external bias[26]. As the current is increased from 0.05 mA to 1.1 mA, the two signals shift 200 and 150 MHz, respectively, to *higher* frequency (Fig. 3b). Moreover, the peak width increases considerably with the current (over a factor of two difference, Fig. 3d). For both peaks, the power is of the order of a few pW and initially increases like $I^2$. The power of the first signal reaches a saturation value at about 0.2 mA (Fig. 3f). Note that for TE-FMR at constant temperature, the power should vary as $I^2$ at best, without taking into account the bias dependence of the resistance.

A similar behaviour is found at -250 Oe and negative current, when both current and field favour the P state (Fig. 2c). For negative applied fields, since the local field is higher in the centre, we attribute the first signal to a precession mode with large amplitude at the ends of the ellipse, and the higher frequency peak to a centre mode. As in the previous case, the peak power varies as $I^2$, reaching a maximum of 12 pW (Fig. 3e). The peak width increases almost three-fold in the considered current range (Fig. 3c). It is noteworthy, however, that for this current-field polarity, the two peaks shift to *lower* frequency with the applied bias. The shift of the ends and centre modes signals amounts to 100 and 180 MHz, respectively (Fig. 3a).

When current and field favour opposite states, the in-plane torque acts against the damping and can bring the magnetization of the free layer into stable precession states. Indeed, we find that under such conditions the sample behaviour is consistently different than obtained for TE-FMR and follows the trends expected for spin-transfer driven precession[3-5,27,28].

At 200 Oe and negative current (Fig. 2a), the field favours the AP state and the spin-torque tends to bring the sample in the P alignment. Below -0.3 mA, the spectra resemble those measured for the same field and similar *positive* bias, though the power increases more rapidly with the current. The two peaks appear at the same frequencies as for positive currents (Fig. 4b; for comparison with TE-FMR frequency, see Supplementary Figure 8), hence the sample remains in the small angle TE-FMR regime. Increasing the current above the switching value (-0.3 mA/-165 mV) induces a gradual red-shift of the signals. The shift - 1.2 GHz (710 MHz) for the ends (centre) signal -, is considerably larger than measured for the opposite current polarity at the same applied field. Both macrospin calculations[3] and full micromagnetic simulations[27,28] attribute such behaviour to an increase of precession angle that occurs when the spin-torque pumps more energy into the system than is lost through damping, rendering higher energy trajectories accessible and effectively driving the system's dynamics. This interpretation is consistent with the significant power increase exhibited by the same signals. Indeed, the output power of the signals measured at 200 Oe and negative current increases noticeably faster than in the TE-FMR regime (Fig. 4d). Most remarkably, the power of the second harmonic of the first signal – which becomes visible at -0.3 mA and dominates the spectra above -0.8 mA - scales roughly as $I^6$, reaching a maximum of 27 nW. The total integrated power is almost four orders of magnitude larger at 1.1 mA (STP) than at -1.1 mA (TE-FMR), and thus cannot be simply attributed to temperature variations in the system. The spectra also develop an increasing $1/f$ tail which is associated with increasingly incoherent precession in the STP regime[27,28].

At -250 Oe and positive current, the field (current) favours the P (AP) state and current-induced precession is again excited (Fig. 2d). The signals attain lower amplitudes than in the case discussed above and the second harmonics remain less important than the main peaks (Fig. 4c). The $1/f$ noise is also lower, suggesting that the dynamics is less chaotic. A red-shift is noted only above 0.7 mA (206 mV) (Fig. 4a; for comparison with TE-FMR frequency, see Supplementary Figure 8). The width of both peaks initially decreases with the current, reflecting the decrease in the effective damping as the spin-torque increases (Supplementary Figure 7). The threshold currents, determined from the extrapolation to zero linewidth, are about 0.36 (0.46) mA for the ends (centre) of the ellipse (Supplementary Discussion).

The precession angle ($\theta_{prec}$) and the tilt angle of the precession axis with respect to the field direction ($\theta_{tilt}$) can be calculated from the power of the fundamental and second harmonic (Supplementary Discussion). Taking that about half of the sample contributes to the lower (higher) frequency signal, $\theta_{prec}$ and $\theta_{tilt}$ at the centre and the ends of the sample are found to vary with the current as shown in Fig. 4e and 4f. In both cases, the precession angle increases linearly with the current, attaining a maximum of 63° at 200 Oe and -1.1 mA (-463 mV). At -250 Oe and 1.1 mA (325 mV), $\theta_{prec}$ is considerably lower (23°), which is consistent with the smaller frequency shift. In all cases, in the low-current region, where the spin-torque is too weak, the data deviate from the linear dependence and saturate at the TE-FMR precession angle, about 5°.

The dynamics of the free layer's magnetization under the influence of a spin-polarized current can be described by a modified Landau-Lifshitz-Gilbert equation[2]:

$$\frac{\partial \vec{m}}{\partial t} = -\gamma \vec{m} \times \vec{H}_{eff} + \alpha \vec{m} \times \frac{\partial \vec{m}}{\partial t} + \frac{\gamma}{M_s Vol} T_{//} \vec{m} \times (\vec{m} \times \vec{m}_r) + \frac{\gamma}{M_s Vol} T_\perp (\vec{m} \times \vec{m}_r) \quad (1)$$

Here, $\vec{m}$ ($\vec{m}_r$) is the unit vector parallel to the magnetization of the free (reference) layer, $\vec{H}_{eff}$ includes the external field as well as any coupling in the system, $Vol$ is the volume of the free layer, $\gamma$ the gyromagnetic ratio and $T_{//}$ and $T_\perp$ are functions of the applied voltage. Note that, unlike $T_{//}$, $T_\perp$ is related to the exchange coupling energy between the two magnetic layers and does not necessarily cancel out when the voltage is set to zero[29]. It has been shown[30] that as long as the spin-torque acts only as a perturbation and the sample remains near static equilibrium, the peak width can be written as the sum of a term given by the intrinsic damping[31], and a second term proportional to the magnitude of the in-plane torque:

$$\Delta f = \frac{\gamma}{2\pi} \alpha (4\pi M_s + 2 H_{eff}) + 2 \frac{\gamma}{2\pi} \frac{T_{//}}{M_s Vol} \quad (2)$$

In the first approximation, the resonance frequency is altered solely by the perpendicular torque:

$$f = \frac{\gamma}{2\pi} \sqrt{\left( H_{eff} + \frac{T_\perp}{M_s Vol} \right) \left( H_{eff} + \frac{T_\perp}{M_s Vol} + 4\pi M_s \right)} \quad (3)$$

It is thus possible to estimate the bias dependence of $T_{//}$ and $T_\perp$ from the voltage variation of the peak width and peak frequency, respectively, when the free layer remains near static equilibrium (in the TE-FMR regime, Fig. 3a-d).

Fig. 5a shows the variation of $T_\perp$ at a given voltage with respect to its zero-voltage value ($\Delta T_\perp/(M_s Vol)$), versus the applied bias, as calculated from the shift of the two TE-FMR peaks (Fig. 3a and 3b). Note that the data for negative bias has been obtained from measurements close to the P state (at -250 Oe), while the positive part comes from TE-FMR signals around the AP alignment (at 200 Oe). (At 200 Oe and positive bias - or -250 and negative bias - , the sample is mainly in the STP regime, where the linear approximation is not valid and this type of analysis cannot be applied). Up to about 300 mV, the general trends for both bias polarities are fitted reasonably well by a quadratic dependence of $T_\perp$ on the voltage, as suggested by theory[17]. The best fit yields $\Delta T_\perp/(M_s Vol) = C_1 * V^2$, with $C_1 = 180$ Oe/V². At 300 mV, a change of regime is found; at higher voltages, the perpendicular torque increases less rapidly with the bias - in a seemingly linear fashion. The values obtained from the two signals at positive bias are in good agreement, as expected, since $\Delta T_\perp/(M_s Vol)$ at a given bias should be independent of the local fields or the excited area. The agreement is not as good at negative bias, probably due to errors in estimating the exact centre frequency of each signal, as the two peaks are very close to each other and the second peak is almost covered by the first signal (Supplementary Figure 5).

The bias dependence of the in-plane spin-torque term $T_\parallel/(M_s Vol)$, calculated from the peak width change in Fig. 3c and 3d is plotted in Fig. 5b. As discussed above, the negative and positive voltage data come from measurements at different field values. The in-plane torque exhibits a linear dependence for low bias (below ±150 mV), and increases more rapidly at higher voltages. The best fits give $T_\parallel/(M_s Vol) = C_2 * V + C_3$, with $C_2 = 24$ Oe/V and $C_3 = 0$ Oe at low voltages, $C_2 = 660$ Oe/V and $C_3 = 90$ Oe for $V < -150$ Oe and $C_2 = 180$ Oe/V and $C_3 = -27$ Oe for $V > 150$ Oe. The larger slope of $T_\parallel$ at high negative voltage - as deduced from the TE-FMR peak width dependence on the bias - is in agreement with the considerably larger power and peak shift measured for this bias polarity in the STP regime (Fig. 2 and Fig 4). Moreover, the in-plane spin-torque change of slope occurs roughly at the threshold currents for STP (discussed above).

From the extrapolation of the peak widths in zero current, the damping constant is estimated to be $\alpha \approx 0.02$ (Supplementary Figure 6), within the reasonable range for samples such as considered here[32,33].

By comparing the data in Fig. 5a and 5b, we find that the perpendicular torque has the same order of magnitude as the in-plane torque, reaching about 25% of its amplitude at the maximum applied bias. This is a particular feature of MTJs, since in GMR samples the perpendicular torque is considerably lower than the in-plane component and can be neglected in most cases[21,22].

The bias dependence of the two spin-torque components at low applied voltage is similar to that determined via ST-FMR experiments[18,19]. Slonczewski and Sun[16] predicted a linear dependence of $T_\parallel$ on voltage for symmetric MTJs or for systems characterized by an asymmetry of elastic tunnelling (with different degrees of dislocation density at the interfaces between the insulating barrier and the two magnetic electrodes). The parabolic variation of $T_\perp$ on $V$ is in good agreement with the theoretical calculations of Theodonis $et\ al.$[17].

The origin of the change of regime demonstrated by the two torque components between 150 and 300 mV is unclear. The change of regime has previously been attributed[19] to an over-estimation of the torque magnitude at currents where heating or hot-electron effects become important, since such effects induce a decrease of the total magnetic moment of the free layer (considered here to be constant). This, however, cannot explain the data in Fig. 5, since $T_\perp$ actually exhibits a slower increase with $V$ above 300 mV. Instead, the change of regime could be linked to an anomaly which appears in the differential tunnelling conductance ($dI/dV$ and $d^2I/d^2V$) spectra of CoFeB/MgO/CoFeB tunnel junctions at similar voltages[34,35]. Indeed, a reduction of the differential conductance $dI/dV$ was found in annealed samples in which the CoFeB electrodes were crystallized in a bcc (001) structure[35]. While the origin of the differential tunnelling conductance anomaly is also unknown, it has been suggested that it may arise from the detailed electronic structure of the interfaces between MgO(001) / bcc CoFeB(001) after annealing, or from a particular feature of the $\Delta_1$ states in MgO(001) or bcc CoFeB(001)[35].

A phase diagram for spin-transfer induced precession can be assembled by plotting the integrated power versus current and field, as shown in Fig. 6. Similar to metallic pillars, the power is high for STP and low when only TE-FMR is measured. When the in-plane spin-torque acts against the intrinsic damping, the power increases with the current (which tends to open the precession cone) and decreases with the field (which tends to reduce the precession angle).

The maximum measured power was 0.14 $\mu W$, but this is estimated to be only about 10% of the real output power of the device, the rest being lost because the sample is not 50 $\Omega$ - adapted and through capacitance effects between current lines (Supplementary Methods). Samples with the same structure but narrower line design provided higher signals (0.43 $\mu W$), though losses are still large. These output power levels are compatible with applications for RF devices. Also important for applications is the stronger bias dependence of the in-plane spin-torque term at negative voltage – which induces a more rapid increase of the power and shift in frequency with the applied bias. The quality factors (Q) of MTJ nano-oscillators are currently below 100 due to their large peak width (~100 MHz). While these values are not satisfactory for any practical purpose, they are comparable to those obtained for GMR pillars in the same field configuration and magnitude[3-5,11]. It should thus be possible to improve Q considerably by employing the same techniques that proved efficient for all-metallic samples, such as increasing the free layer's volume, phase-locking on another oscillator[7,8], or changing the angle and magnitude of the applied field[11].

1. Slonczewski, J.C., Current-driven excitations of magnetic multilayers. *J. Magn. Magn. Mater.* **159**, L1-L7 (1996).

2. Berger, L., Emission of spin waves by a magnetic multilayer traversed by a current. *Phys. Rev. B* **54,** 9353-9358 (1996).

3. Kiselev, S.I. *et al.*, Microwave oscillations of a nanomagnet driven by a spin-polarized current. *Nature* **425**, 380-382 (2003).

4. Covington, M., AlHajDarwish, M., Ding, Y., Gokemeijer, N.J. and Seigler, M.A., Current-induced magnetization dynamics in current perpendicular to the plane spin valves. *Phys. Rev. B* **69**, 184406 (2004).

5. Deac, A., *et al.*, Study of spin-transfer-induced dynamics in spin-valves for current-perpendicular-to-plane magnetoresistive heads. *J. Phys.: Condens. Matter* **19**, 165208 (2007).

6. Krivorotov, I.N. *et al.*, Time-domain measurements of nanomagnet dynamics driven by spin-transfer torques. *Science* **307**, 228-231 (2005).

7. Kaka, S. *et al.*, Mutual phase-locking of microwave spin-torque nano-oscillators. *Nature* **437**, 389-392 (2005).

8. Mancoff, F.B., Rizzo, N.D., Engel, B.N. and Tehrani, S., Phase-locking in double-point-contact spin-transfer devices. *Nature* **437**, 393-395 (2005).

9. Tulapurkar, A.A. *et al.*, Spin-torque diode effect in magnetic tunnel junctions. *Nature* **438**, 339-342 (2005).

10. Sankey, J.C. *et al.*, Spin-transfer driven ferromagnetic resonance of individual nanomagnets. *Phys. Rev. Lett.* **96**, 227601 (2006).

11. Rippard, W.H., Puffal, M.R., Kaka, S., Silva, T.J. and Russek, S.E., Current-driven microwave dynamics in magnetic point contacts as a function of applied field angle. *Phys. Rev.B* **70**, 100406R (2004).

12. Yuasa, S., Nagahama, T., Fukushima, A., Suzuki, Y. and Ando, K., Giant room-temperature magnetoresistance in single-crystal Fe/MgO/Fe magnetic tunnel junctions. *Nature Mater.* **3**, 868 (2004).

13. Parkin, S.S.P. *et al.*, Giant tunnelling magnetoresistance at room temperature with MgO (100) barriers. *Nature Mater.* **3**, 862 (2004).

14. Yuasa, S., Fukushima, A., Kubota, H. and Ando, K., Giant tunneling magnetoresistance up to 410 % at room temperature in fully epitaxial Co/MgO/Co tunnel junctions with bcc Co(001) electrodes. *Appl. Phys. Lett.* **89**, 042505 (2006).

15. Hayakawa, J., Ikeda, S., Lee, Y.M., Matsukura, F. and Ohno, H., Effect of high annealing temperature on giant tunnel magnetoresistance ratio of CoFeB/MgO/CoFeB magnetic tunnel junctions. *Appl. Phys. Lett.* **89**, 232510 (2006).

16. Slonczewski, J.C. and Sun, J., Theory of voltage-driven current and torque in magnetic tunnel junctions. *J. Magn. Magn. Mater.* **310**, 169-175 (2007).

17. Theodonis, I., Kioussis, N., Kalitsov, A., Chshiev, M. and Butler, W.H., Anomalous bias dependence of spin-torque in magnetic tunnel junctions. *Phys. Rev. Lett.* **97**, 237206 (2006).

18. Kubota, H. *et al.*, Quantitative measurement of voltage dependence of spin-transfer torque in MgO-based magnetic tunnel juntions. *Nature Phys.* **4**, 37 (2008).

19. Sankey, J.K. *et al.*, Measurement of the spin-transfer-torque vector in magnetic tunnel juntions. *Nature Phys.* **4**, 67 (2008).

20. Nazarov, A.V. *et al.*, Spin transfer stimulated microwave emission in MgO magnetic tunnel junctions. *Appl. Phys. Lett.* **88**, 162504 (2006).

21. Myers, E.B. *et al.*, Thermally activated magnetic reversal induced by a spin-polarized current. *Phys. Rev. Lett.* **89**, 196801 (2002).

22. Zimmler, M.A. *et al.*, Current-induced effective magnetic fields in Co/Cu/Co nanopillars. *Phys. Rev. B* **70**, 184438 (2004).

23. Kubota, H. *et al.*, Dependence of spin-transfer switching current on free layer thickness in Co-Fe-B/MgO/Co-Fe-B magnetic tunnel junctions. *Appl. Phys. Lett.* **89**, 032505 (2006).

24. Tsunekawa, K. *et al.*, Giant tunnelling magnetoresistance effect in low-resistance CoFeB/MgO(001)/CoFeB magnetic tunnel junctions for read-head applications. *Appl. Phys. Lett.* **87**, 072503 (2005).

25. Kittel, C., *Introduction to Solid State Physics*. $7^{th}$ edition, p. 505 (Wiley, New-York, 1996).

26. McMichael, R.D. & Stiles, M.D, Magnetic normal modes of nanoelements. *J. Appl. Phys.* **97**, 10J901 (2005).

27. Lee, K.-J., Deac, A., Redon, O, Nozieres, J.-P. and Dieny, B., Excitations of incoherent spin-waves due to spin-transfer torque. *Nature Mater.* **3**, 877 (2004).

28. Zhu, J.G. and Zhu, X.C., Spin transfer induced noise in CPP read heads. *IEEE Trans. Magn.* **40**, 182-188 (2004).

29. Slonczewski, J.C., Currents, torques, and polarization factors in magnetic tunnel junctions. *Phys. Rev. B* **71**, 024411 (2005).

30. Petit, S. *et al.*, Spin-torque influence on the high-frequency magnetization fluctuations in magnetic tunnel junctions. *Phys. Rev. Lett.* **98**, 077203 (2007).

31. Zhu, J.G., Thermal magnetic noise and spectra in spin valve heads. *J. Appl. Phys.* **91**, 7273 (2002).

32. Fuchs, G.D. *et al.*, Spin-torque ferromagnetic resonance measurements of damping in nanomagnets. *Appl. Phys. Lett.* **91**, 062507 (2007).

33. Bilzer, C., Study of dynamic magnetic properties of soft CoFeB films. *J. Appl. Phys.* **100**, 053903 (2006).

34. Matsumoto, R. *et al.*, Tunneling spectra of sputter-deposited CoFeB/MgO/CoFeB magnetic tunnel junctions showing giant tunneling magnetoresistance effect. *Solid State Comm.* **136**, 611 (2005).

35. Matsumoto, R. *et al.*, Dependence on annealing temperatures of tunneling spectra in high-resistance CoFeB/MgO/CoFeB magnetic tunnel junctions. *Solid State Comm.*, **143**, 574 (2007).

**Supplementary Information** accompanies the paper.

**Acknowledgements** The authors thank J.C. Slonczewski, G.E.W. Bauer, T.J. Silva, S.E. Russek and M. Mizuguchi for discussions. This research was partially supported by SCOPE-MIC and 21COE-MEXT. A.D. acknowledges support from the Japan Society for the Promotion of Science through the research grant P05330.

**Authors' Contributions** A.D. carried out the measurements, analyzed the data and wrote the paper. A.F. and H.K. developed the patterning process and fabricated the samples. A.F. also provided considerable help with the experimental setup. H.K., Y.N., K.T., D.D. and D.W. optimised the sputtering process for low-RA MgO magnetic tunnel junctions and deposited the multilayer. S.Y. helped develop the magnetic tunnel junction and carried on studies for increasing the TMR. Y.S. helped with analysing the data and calculated the losses in the system. All authors discussed the result and commented on the manuscript.

**Competing interests statement** The authors declare that they have no competing financial interests.

**Correspondence** and requests for materials should be addressed to A.D. (alina.deac@excite.com).

**Figure 1**: Sample properties and principle of the experiment. **a.** Sample structure, measurement principle and sign convention for current and field. Precession states excited in the free layer of a MTJ generate an oscillatory voltage which is monitored on a spectrum analyzer. Positive fields favour the AP state; for positive current, the electrons flow from the free to the reference layer, also favouring the AP alignment. **b.** Differential resistance as function of magnetic field. (Inset) Zoom-in in the low-field range.

**Figure 2**: Spectra measured at 200 and -250 Oe and positive (negative) currents between 0.1 (-0.1) and 1.1 (-1.1) mA with 0.1 mA steps. On each graph, the spectra have been progressively shifted upwards with 2.5 nV/Hz$^{1/2}$ for increasing currents. **a.** Microwave signals measured for 200 Oe and negative currents. **b.** Spectra obtained at 200 Oe and positive currents. **c.** Microwave signals for -250 Oe and negative currents. **d.** Spectra recorded at -250 Oe and positive currents. When current and field favour the same state, only TE-FMR signals are obtained (**b** and **c**). Current-induced stable precession states appear if current and field favour opposite alignments (**a** and **d**).

**Figure 3**: TE-FMR: frequency shift with the current, peak width dependence on the current and integrated power of the main peaks versus the square of the applied current. The data has been extracted from the spectra in Figure 2b and 2c. **a** and **b.** Frequency current-dependence for the two main peaks at -250 and 200 Oe, respectively, in the TE-FMR regimes. Squares and triangles correspond to signals obtained from the ends and the centre of the sample, respectively. **c** and **d.** Peak width versus current, at -250 and 200 Oe. In red: data points corresponding to the ends mode. In black: data for the centre mode. The peak width of the ends mode at -250 Oe cannot be precisely determined, as these signals are almost completely covered by the peak attributed to centre

mode. . **e** and **f.** Integrated power for fundamental signals at -250 and 200 Oe, as a function of the square of the applied current. Black (red) squares: fundamental power for the ellipse ends (centre).

**Figure 4**: Current-induced precession: current-dependence of peak frequency, power, precession and tilt angle dependence at -250 and 200 Oe. The data has been extracted from the spectra in Figure 2a and 2d. **a** and **b.** Current-dependence of frequency for the two main peaks at -250 and 200 Oe, respectively, in the current-induced precession regimes. Squares and triangles correspond to signals obtained from the ends and the centre of the sample, respectively. **c** and **d.** Integrated power for the two main peaks and their second harmonics at -250 and 200 Oe, as a function of the square of the applied current. Black (red) squares: fundamental power for the ellipse ends (centre); black (red) triangles: power of the second harmonics of the same signals. **e** and **f.** Current-dependence of precession and tilt angle, at -250 and 200 Oe. Open (solid) black squares: precession angle at the centre (ends); open (solid) red circles: tilt angle at the centre (ends). In blue: linear fits for the precession angle.

**Figure 5**: Parallel and perpendicular torque dependence on the bias. The voltage has been calculated from the applied current and the resistance variation with the current, in the P and AP state, respectively. **a.** Variation of the perpendicular spin-transfer torque term with respect to its zero-voltage value, versus bias, as estimated from the TE-FMR peak shift with the voltage (see Fig. 3a and 3b). **b.** Parallel spin-transfer torque component versus bias, calculated from the TE-FMR peak width change with the voltage (see Fig. 3c and 3d), as explained in the text. In red: values obtained from the ends mode data. In black: values deduced from the centre mode data. Full (empty) symbols mark data points obtained from measurements close to the P (AP) state. Blue lines: best fit

(as explained in the text). The voltage has been estimated from the current dependence of the resistance in the two states.

**Figure 6**: Current-field phase diagram obtained by plotting the power integrated over all the measured frequency range for each current and field. The colour scale marks the attained power levels. Each graph in Fig. 2 corresponds to one quadrant of the phase diagram.

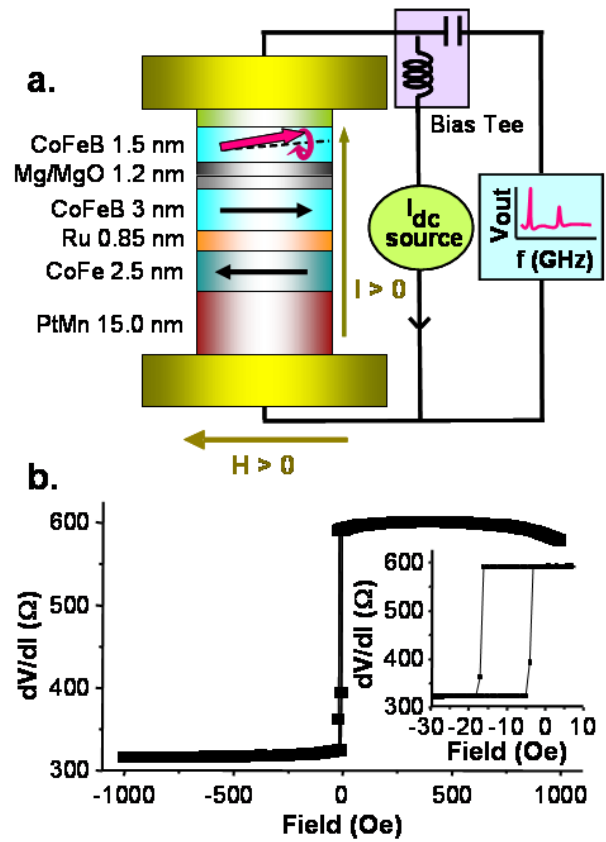

**Figure 1**

Deac et al. _ Bias-driven large power microwave emission from MgO-based tunnel magnetoresistance devices

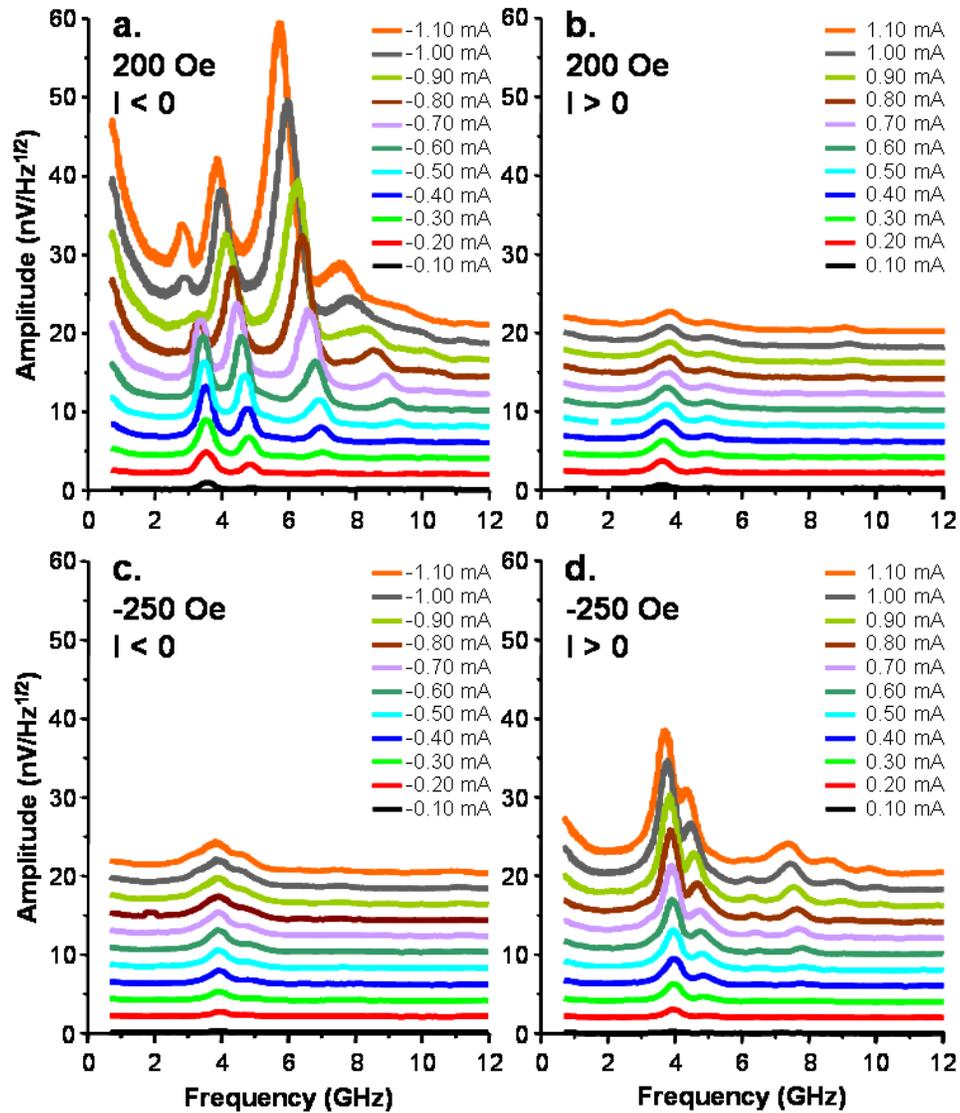

**Figure 2**

Deac et al. _ Bias-driven large power microwave emission from MgO-based tunnel magnetoresistance devices

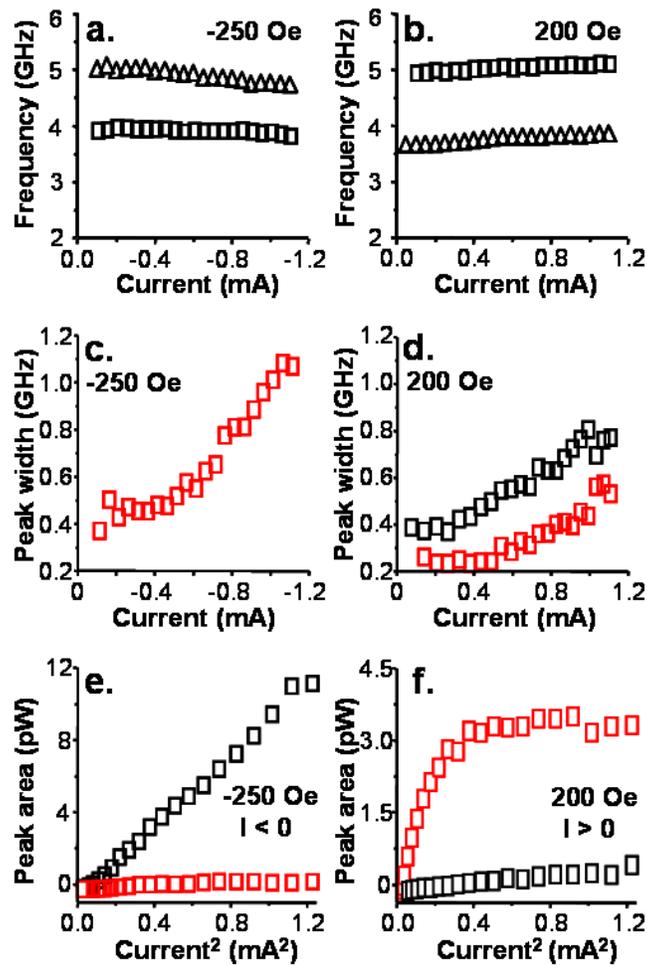

**Figure 3**

Deac et al. _ Bias-driven large power microwave emission from MgO-based tunnel magnetoresistance devices

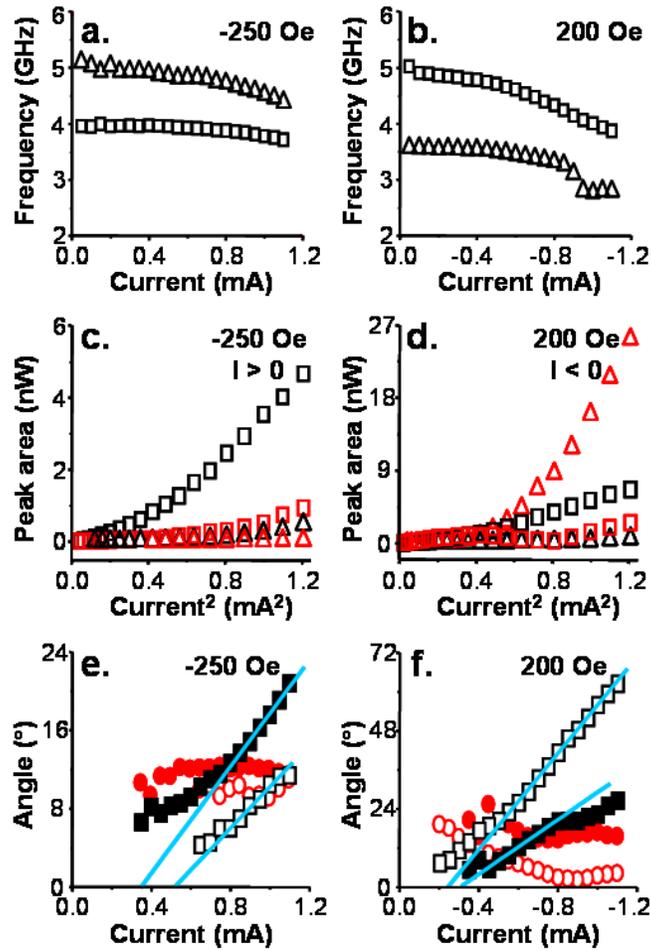

**Figure 4**

Deac et al. _ Bias-driven large power microwave emission from MgO-based tunnel magnetoresistance devices

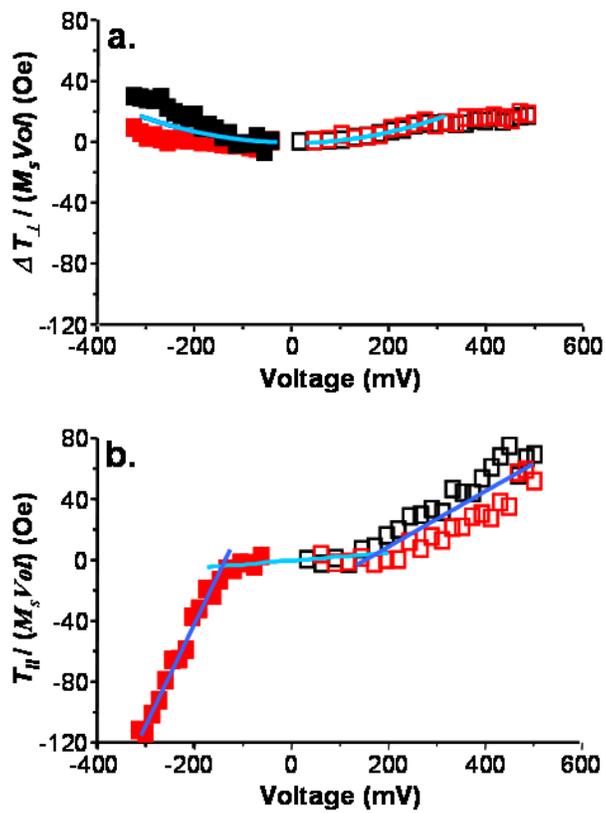

**Figure 5**

Deac et al. _ Bias-driven large power microwave emission from MgO-based tunnel magnetoresistance devices

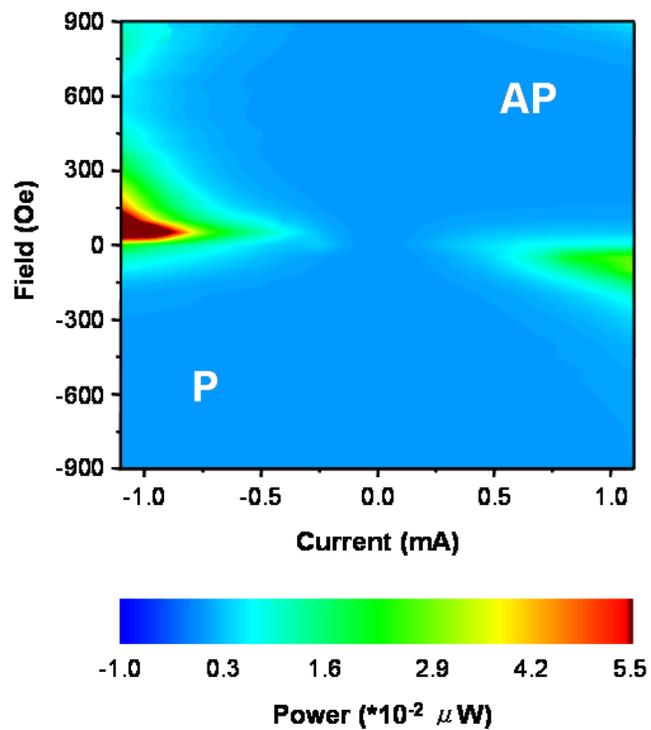

**Figure 6**

Deac et al. _ Bias-driven large power microwave emission from MgO-based tunnel magnetoresistance devices

# Bias-driven large power microwave emission from MgO-based tunnel magnetoresistance devices

Alina M. Deac, Akio Fukushima, Hitoshi Kubota, Hiroki Maehara, Yoshishige Suzuki, Shinji Yuasa, Yoshinori Nagamine, Koji Tsunekawa, David D. Djayaprawira and Naoki Watanabe

**Supplementary information**

**Supplementary methods:**

The sample fabrication procedure plus a detail static characterization of pillars such as analyzed here has been presented elsewhere (see ref. 23 and 24 in the main text).

**1. Switching**

Resistance versus current and magnetoresistance curves were measured using a four-point method. The setup allowed both for dc resistance ($R$) and for differential resistance ($dV/dI$) measurements with a lock-in technique. Differential resistance versus field curves were measured with 1 μA, 9.7 kHz ac current.

Supplementary Fig. 1 shows the current dependence of the *resistance* in the P ($R_P$) and AP ($R_{AP}$) states for the sample discussed in the text. The data has been extracted from $R(H)$ curves measured for different dc currents. The dc bias was varied from -1.1 to 1.1 mA, with a 0.05 mA step (the same as for high frequency measurements).The resistance of the AP state was 599 Ω in zero bias and showed a strong, approximately linear, decrease with the current. The decrease was sharper for negative than for positive current: at -1.1 mA, $R_{AP}$ was as low as 421 Ω, versus 454 Ω at +1.1 mA. The resistance of the parallel state was roughly constant. The slight linear decrease of $R_P$ is very likely due to heating, since it follows the sense of the current variation. The magnetoresistance dropped from 110% at zero bias to 49% and 53% at -1.1 and +1.1 mA, respectively.

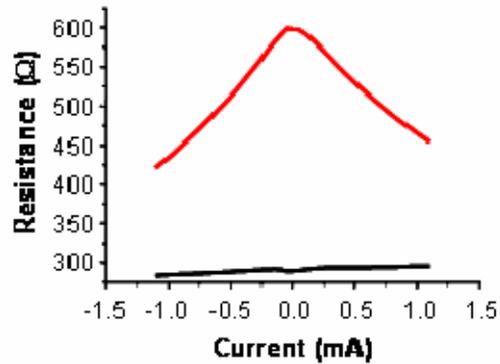

**Supplementary Figure 1:** Current dependence of the P and AP state resistance as determined from magnetoresistance curves measured at different dc currents. In red (black): the resistance of the sample in the AP (P) state.

In order to avoid over-heating the sample, *differential resistance* versus current curves were measured by applying consecutive current pulses with increasing/decreasing amplitude. The current pulse duration was 100 ms, followed by a waiting time of 400 ms before the next pulse was applied. The differential resistance was measured both on-pulse and during the waiting time (off-pulse). The on-pulse curves reflect the bias dependence of the differential resistance, as well as peaks or noise in the current/field range where RF excitations are obtained. The off-pulse measurements record only switching events which may have occurred.

Supplementary Fig. 2 shows *dV/dI* versus current curves, measured off-pulse and on-pulse, in an external field that compensates the shift of the zero-dc current magnetoresistance loop corresponding to the switching of the free layer. The switching currents are about +/-0.3 mA. Since the field distribution inside the free layer is inhomogeneous and the samples are fairly susceptible to thermal activation, these values are to be taken as order of magnitude rather than absolute switching currents.

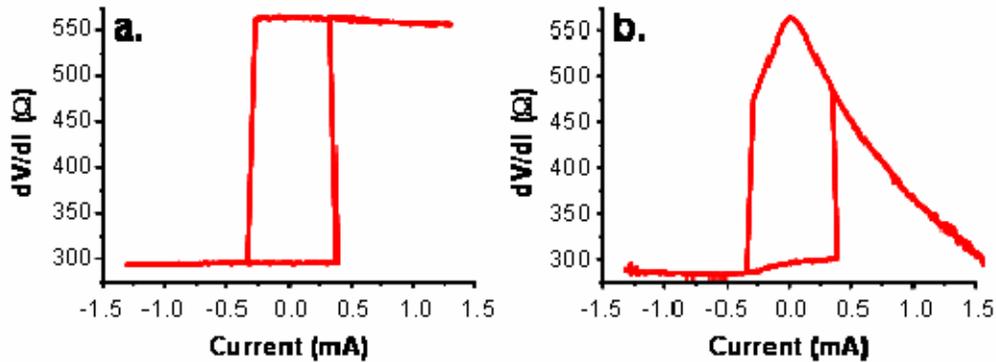

**Supplementary Figure 2:** Resistance versus current measured (**a.**) off-pulse and (**b.**) on-pulse, under an applied field that compensates the shift of the minor magnetoresistance loop (corresponding to the switching of the free layer between the parallel and antiparallel state).

Microwave measurements were performed by sweeping the current at a fixed field. After the field was set, the current was ramped up to -1.1 mA. Spectra were then measured for each current value between -1.1 and 1.1 mA, after which the current was set back to zero, the field was changed and the procedure repeated.

Note that while all RF experiments were conducted at constant current, so as to insure a better protection for the sample, the relevant parameter for tunnel junctions is actually the voltage. All through the text, the current was converted into voltage by taking into account the resistance variation with the current.

## 2. High frequency measurements: modelling for loss evaluation and peak power estimation.

For high frequency experiments, power losses appear in the system because the sample is not 50 Ω - adapted, because of capacitance effects between the top and bottom current lines (which are several tens of micrometers wide in the area where they superpose and connect to the sample), as well as capacitive coupling through the silicon substrate. To evaluate the effect of parasitic impedances, the measurement system is modelled as seen in Supplementary Fig. 3a. In the figure, $R_T$ represents the resistance of the magnetic tunnel junction, which is connected to the 50 Ω input of a spectrum analyzer through a series parasitic impedance $Z_s$, a 50 Ω RF wave guide and the capacitance of the bias-Tee. A constant current is applied to the tunnel junction through the inductor in the bias-Tee. The parasitic impedance of $Z_p$ is in parallel with the tunnelling resistance. In our case, $Z_p$ is composed of two capacitors with certain losses (Supplementary Fig. 3b). $C_{p1}$ and $R_{p1}$ model capacitance effects and losses between the upper and the lower current lines (electrodes) around the MTJ. The capacitive coupling between the electric contact pads through the silicon substrate is modelled via $C_{p2}$ and $R_{p2}$. $Z_s$ can be considered to be a pure inductor.

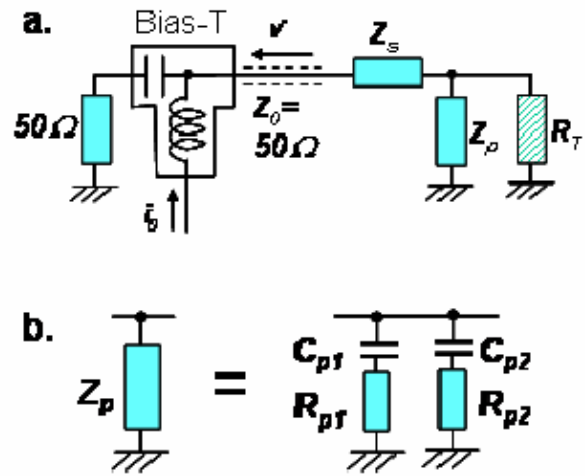

**Supplementary Figure 3:** Model of the sample in the measurement system. **a.** Model of the entire system. **b.** The parasitic impedance $Z_p$ is composed of two capacitors with certain losses.

The MTJ itself is modelled as an oscillating resistance. When precession states are excited in the free layer, the angle between its magnetization and that of the pinned layer oscillates around an average tilt angle $\theta_{tilt}$ with an angular frequency $\omega_0$:

$$\theta = \theta_{tilt} + \theta_{prec} \sin(\omega_0 t)$$

where $\theta_{prec}$ is the precession angle. Consequently, the resistance of the tunnel junction oscillates as:

$$R_T = \frac{R_{AP} + R_P}{2} - \frac{R_{AP} - R_P}{2} \cos\left(\theta_{tilt} + \theta_{prec}\sin(\omega_0 t)\right)$$
$$= \left(\frac{R_{AP} + R_P}{2} - \frac{R_{AP} - R_P}{2} J_0(\theta_{prec})\cos\theta_{tilt}\right) +$$
$$+ (R_{AP} - R_P)\left(J_1(\theta_{prec})\sin\theta_{tilt}\sin(\omega_0 t) - J_2(\theta_{prec})\cos\theta_{tilt}\cos(2\omega_0 t) + \ldots\right)$$
$$\equiv R_0 + \left(\frac{\Delta R_1}{\sqrt{2}} e^{i\omega_0 t} + c.c.\right) + \left(\frac{\Delta R_2}{\sqrt{2}} e^{2i\omega_0 t} + c.c.\right) + \ldots,$$

where $R_0$, $|\Delta R_1|$, and $|\Delta R_2|$ are defined as follows:

$$\begin{cases} R_0 \equiv \dfrac{R_{AP} + R_P}{2} - \dfrac{R_{AP} - R_P}{2} J_0(\theta_{prec})\cos\theta_{tilt} \\ |\Delta R_1| \equiv \dfrac{1}{\sqrt{2}}(R_{AP} - R_P)J_1(\theta_{prec})\sin\theta_{tilt} \\ |\Delta R_2| \equiv \dfrac{1}{\sqrt{2}}(R_{AP} - R_P)J_2(\theta_{prec})\cos\theta_{tilt} \end{cases}.$$

Taking into account that the ω-component of the voltage across the MTJ, $v_T$, is given by the following equation:

$$v_T(\omega) = i_0 \Delta R(\omega) + R_0 i_T(\omega),$$

we obtain the outgoing voltage wave amplitude in the waveguide, $v^-$:

$$v^-(\omega) = \frac{2Z_{/\!/}(\omega)}{Z(\omega) + Z_0} \frac{Z_0}{R_0} \frac{\Delta R(\omega)}{2} i_0$$

Where

$$\begin{cases} Z_{/\!/}^{-1}(\omega) = R_T^{-1} + Z_p^{-1} \\ Z(\omega) = Z_{/\!/}(\omega) + Z_s(\omega) \end{cases}.$$

Therefore, the RF output power can be written as:

$$P(\omega) = \frac{|v^-|^2}{Z_0} = \eta(\omega)\left|\frac{\Delta R(\omega)}{R_0}\right|^2 \frac{R_0 i_0^2}{4}.$$

$\eta(\omega) \equiv \dfrac{Z_0}{R_0}\left|\dfrac{2Z_{//}(\omega)}{Z(\omega)+Z_0}\right|^2$ expresses the efficiency of the RF circuit at frequency $\omega$ and is unity when $Z_p \rightarrow$ infinity, $Z_s = 0$ and $R_T = Z_0$. Replacing $\Delta R$ by the above expression, we finally get the following formulas to estimate the output power for the fundamental signal and its corresponding second harmonic:

$$P(\omega_0) = \eta(\omega_0)\left(\frac{R_{AP}-R_P}{R_0}\right)^2 J_1^2(\theta_{prec})\sin^2\theta_{tilt}\frac{R_0 i_0^2}{8},$$

$$P(2\omega_0) = \eta(2\omega_0)\left(\frac{R_{AP}-R_P}{R_0}\right)^2 J_2^2(\theta_{prec})\cos^2\theta_{tilt}\frac{R_0 i_0^2}{8}. \quad (S1)$$

For small precession angles $\theta_{prec}$, the power for the fundamental and the second harmonics can be approximated as:

$$P(\omega_0) = \eta(\omega_0)\left(\frac{R_{AP}-R_P}{R_0}\right)^2 (\theta_{prec})^2 \sin^2\theta_{tilt}\frac{R_0 i_0^2}{32},$$

$$P(2\omega_0) = \eta(2\omega_0)\left(\frac{R_{AP}-R_P}{R_0}\right)^2 \left(\frac{\theta_{prec}}{2}\right)^4 \cos^2\theta_{tilt}\frac{R_0 i_0^2}{32}.$$

In our case, $\eta(\omega)$ was obtained from the fitting of the measured microwave reflective amplitude spectrum, $S_{11}$, based on the model circuit described above. $S_{11}$ was measured on the sample described in the text, in the P and AP states, using a network analyzer. Considering $R_T = 650\ \Omega$, the parasitic resistances/capacitances are found to be: $C_{p1} = 0.6$ pF, $R_{p1} = 11\ \Omega$, $C_{p2} = 2.0$ pF, $R_{p2} = 300\ \Omega$, $Z_s = 12$ nH. Similar calculations were performed for the parallel alignment case as well.

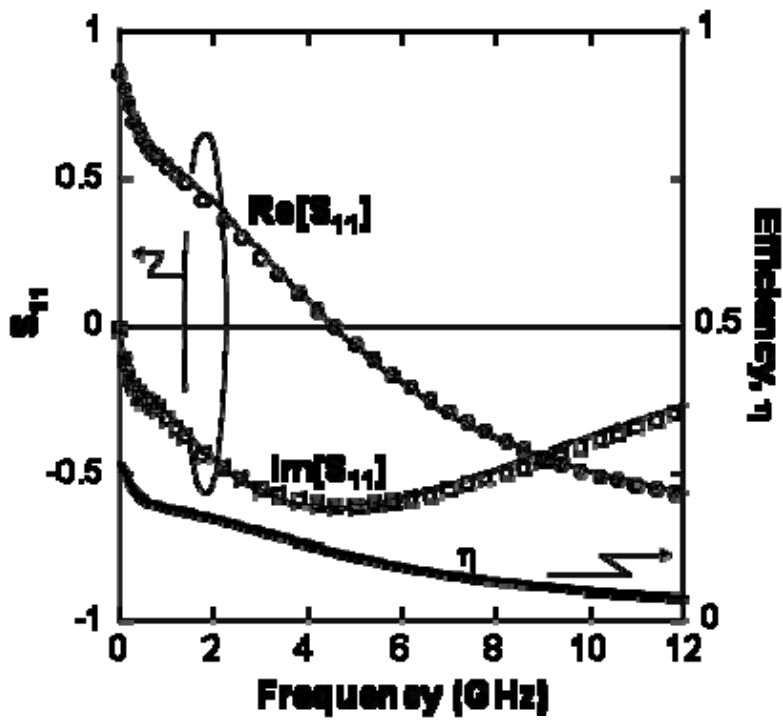

**Supplementary Figure 4:** Measured and calculated reflective amplitude $S_{11}$ spectrum for the antiparallel configuration. The circles (squares) mark the measured data for the real (imaginary) part. The fine lines connecting the data mark the fitting based on the model described above, with the parameters values given in the text. The bold line depicts the RF efficiency of the circuit, as function of frequency, calculated from the same model, with the same parameters.

**Supplementary discussion:**

**1. TE-FMR**

The FMR frequency, in the absence of any coupling fields, can be calculated using Kittel's formula[25]:

$$f = \frac{-\gamma}{2\pi}\sqrt{(H+H_k)(H+H_k+4\pi M_s)}$$

where $\gamma = -1.76*10^{11}$ s$^{-1}$T$^{-1}$ is the gyromagnetic ratio, $H$ is the applied field, $H_k$ is the uniaxial anisotropy field and $M_s$ is the saturation magnetization.

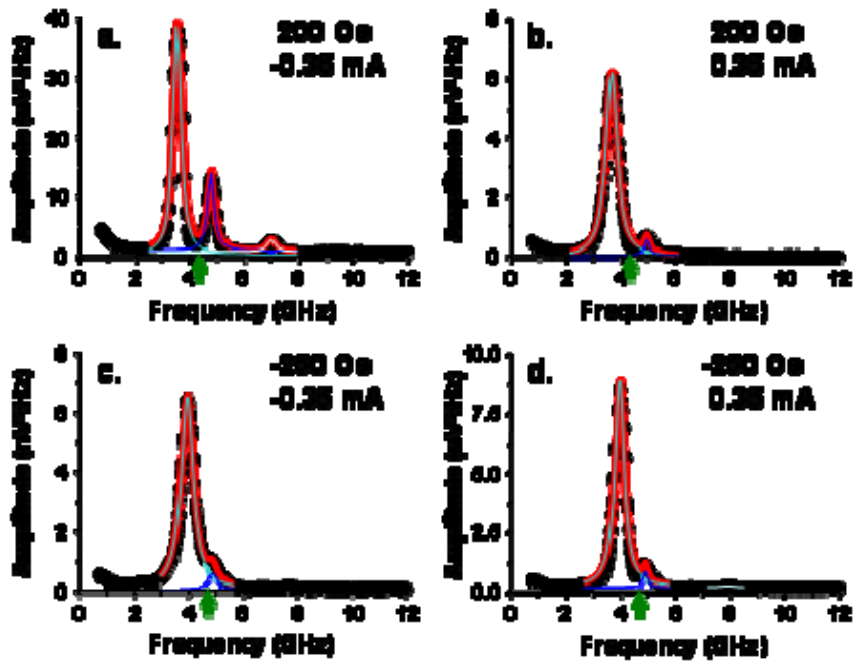

**Supplementary Figure 5:** Spectra in the TE-FMR precession regime, at 200 Oe (**a.** and **b.**) and at -250 Oe (**c.** and **d.**), for positive and negative currents. The black dots mark the measured data, the light and dark blue lines are Lorentzian fits to the two main peaks and the red lines are sums of the two Lorentzians. Also visible on the spectra in **a.** and **d.** is the second harmonic of the first peak. The green arrows mark the precession frequency as calculated using Kittel's formula.

Taking the measured values $H_k = 6$ Oe and $M_s = 880$ emu/cm$^3$, this formula predicts that the precession frequency in the TE-FMR regime should be around 4.26 and 4.77 GHz at 200 and -250 Oe, respectively. The two main peaks on the spectra measured in the TE-FMR precession regime fall on each side of the calculated precession frequency (Supplementary Fig. 5). We thus attribute them to a centre and an ends mode, as explained in the text. Note that centre and ends modes are among the natural precession modes for an elliptic magnetic particle such as the free layer[26].

## 1a. Coupling fields

The coupling fields between the free and the reference layer depend on the exact micromagnetic configuration of the sample in a given external field. They include the magnetostatic interaction between the two layers (favouring the AP state), the Neel coupling through barrier roughness (favouring the P sate) and the contribution of the perpendicular spin-torque term, which is related to the exchange energy, and whose magnitude depends on the exchange splitting[16,17,29]. The average coupling fields at the centre and the long ends of the ellipse can be estimated using the Kittel formula[25], as the additional fields necessary to fit the zero-current extrapolation of the centre or ends peak frequency in the TE-FMR regime, at a given external field:

$$f = \frac{-\gamma}{2\pi}\sqrt{(H + H_k + H_{coupling})(H + H_k + +H_{coupling} + 4\pi M_s)}$$

For both external fields considered here, we find that the average coupling field is positive (favouring the AP state) at the edges of the ellipse and negative (favouring the P alignment) in the centre:

| External field (Oe) | Coupling field ends (Oe) | Coupling field centre (Oe) |
|---|---|---|
| -250 | 80 | -20 |
| 200 | 70 | -55 |

**Supplementary Table 1:** Coupling fields at the ends and the centre of the free layer, for applied field values of -250 Oe and 200 Oe.

**1b. Peak shift with the current in the TE-FMR precession regime; bias dependence of the perpendicular spin-torque term**

While in the TE-FMR precession regime, applying an increasing current through the sample may induce a shift in frequency via three different mechanisms: heating, Oersted fields and field-like spin-torque term ($T_\perp$).

While it is reasonable to assume that there is a certain increase in temperature with the current, heating cannot explain the peak shift observed for the sample described here. Indeed, it has been shown that increasing the temperature induces a *red-shift* of the precession frequency[36]. While the two peaks measured at -250 Oe do undeniably shift to lower frequency as the (negative) current is increased (Fig. 3a), we find that at 200 Oe and positive current (Fig. 3b) the peaks shift to *higher* frequency with increasing bias. This would imply that the sample temperature is actually decreasing at higher currents, which seems unreasonable.

If the peak shift in the TE-FMR precession regime is caused by a current-dependent magnetic field (either Oersted field or perpendicular spin-torque term), the magnitude of this field can be estimated from the precession frequency measured at a given bias using the Kittel formula. The current-induced field values thus calculated are shown in Fig. 5a. Given the remarkable agreement between the values obtained for the centre and ends of the ellipse at positive bias (where the two peaks are well defined), we discard the Oersted field as possible cause for the peak shift, since Oersted fields should show a considerable spatial variation. Moreover, an Oersted field should vary linearly with the bias, not qudraticaly, as seen in Fig. 5a. We thus interpret the peak shift measured in the TE-FMR precession regime as primarily reflecting the bias dependence of $T_\perp$.

**1c. Damping**

Since the in-plane spin-torque term cancels out when the current is turned off, the intrinsic damping constant ($\alpha$) characterizing the dynamics of the free layer can be evaluated using equation (2) in the main text, using the zero-current extrapolation of the width of the two signals. Considering the coupling fields in Supplementary Table 1 and the peak width dependence on the current in the TE-FMR precession regime (Fig. 3c and 3d), the damping constant is estimated to be approximately 0.02 (Supplementary Fig. 6).

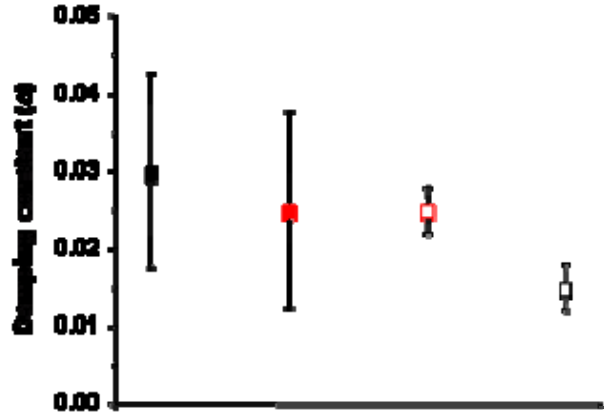

**Supplementary Figure 6:** Damping constant (α), as calculated from the extrapolation in zero-current of the width of the centre (black) and ends (red) TE-FMR peaks at 200 (empty symbols) and -250 Oe (full symbols).

### 1d. Peak width versus current in the TE-FMR precession regime

When current and field favour opposite states, the sample remains in the TE-FMR precession regime as long as the bias remains below the threshold for STP. At higher currents, the magnetization dynamics is driven by the spin-torque. However, even below the threshold, the in-plane spin-torque term opposes the intrinsic damping and should induce a decrease of the TE-FMR peak width with increasing bias. The threshold currents, defined as the bias where $T_{//}$ effectively compensates the damping and steady-state precession is obtained, can be determined as the current values where the peak width extrapolates to zero. At -250 Oe, the threshold currents are estimated to be 0.38 mA for the ends and 0.46 mA for the centre of the free layer (Supplementary Fig. 7). The threshold currents for the opposite current/field polarity (200 Oe and negative current) cannot be determined from the peak width dependence on the bias, since the sample enters the non-linear precession regime at comparatively lower currents, as indicated by the strong peak frequency shift (Supplementary Fig. 8). Additionally, at this current/field polarity, in the STP regime, the dynamics becomes strongly incoherent at relatively low currents, leading to peak widening and amplitude loss, as well as increasing *1/f* noise. Under such circumstances, the peak width is not a measure of the effective damping anylonger[27,28]. As seen in Fig. 2a and 2d, such effects are considerably stronger at 200 Oe and negative bias, than at -250 Oe and positive values, at similar current amplitudes.

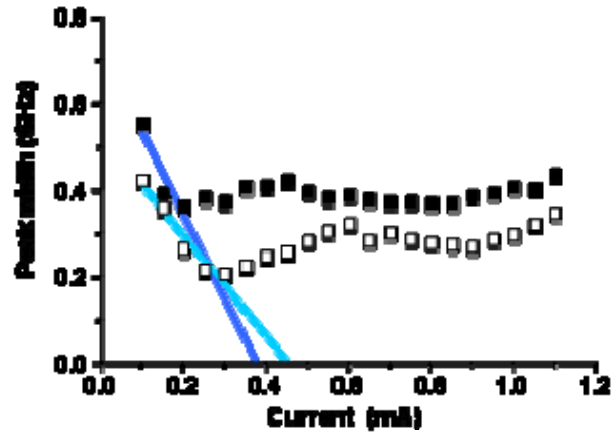

**Supplementary Figure 7:** Peak width versus current at -250 Oe (favouring the P state) and positive current (favouring the AP state), for the centre (empty symbols) and ends signals (full symbols).

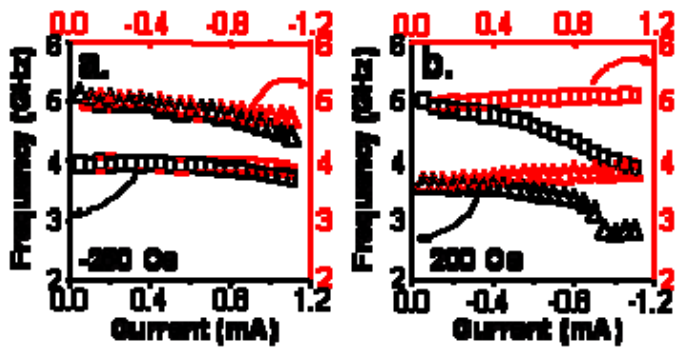

**Supplementary Figure 8:** Comparison between signal frequency in the TE-FMR and STP regimes, at -250 and 200 Oe. The dynamics becomes non-linear when the precession frequency deviates from the TE-FMR values. **a** and **b.** Current-dependence of frequency for the two main peaks at -250 and 200 Oe, respectively. In black: STP signals. In red: TE-FMR peak frequencies, at the same fields. Red and black data points were obtained for currents of the same amplitude but different signs. Squares and triangles correspond to signals obtained from the ends and the centre of the sample, respectively.

## 2. Current dependence of the precession and tilt angles in the spin-transfer driven precession regime

The linear dependence of the precession angle on the applied current is empirical and based on the assumption that the area of the free layer which contributes to each of the two main signals is constant when increasing the current, which may not be realistic. If the linear dependence has a physical meaning, the intersection of the linear fits with the current axis should give the threshold current for spin-transfer induced precession. At -250 Oe and positive currents, the linear fits yield 0.36 mA for the ends mode and 0.52 mA for the centre signal (Fig. 4e). These values are close to the threshold currents obtained from the peak width bias dependence – 0.38 mA for the ends and 0.46 mA for the centre (Supplementary Fig. 7) -, and are therefore reasonable. At 200 Oe and negative current, the currents where the precession angle extrapolates to zero are around -0.24 mA at the centre and -0.32 mA for the ends of the ellipse (Fig. 4e).

The tilt angle can be as high as 25° at 200 Oe and around 12° at -250 Oe, where $\theta_{tilt}$ at the ends and the centre of the ellipse also appears to have more homogeneous values. Accordingly, at -250 Oe the dynamics should be more coherent, in agreement with the lower measured $1/f$ noise (note that the calculation of the tilt and precession angle is based only on the power of the fundamental and second harmonics and does not take into account the $1/f$ noise). Possibly the dynamics is less coherent and attains higher angles at positive fields due to the fact that in this configuration the external field opposes the ferromagnetic coupling and induces a less homogeneous initial micromagnetic state, with larger differences in the tilt angle at the centre and ends of the sample.

**Supplementary notes:**

36. Stutzke, N., Burkett, S.L. and Russek, S.E., Temperature and fields dependence of high-frequency magnetic noise in spin-valve devices. *Appl. Phys. Lett.* **82**, 91 (2003).